\documentclass[%
superscriptaddress,
preprint,
 amsmath,amssymb,
aip,apl,
showkeys,
]{revtex4-2}
\usepackage[dvipdfmx]{graphicx}
\usepackage[dvipdfmx]{color}
\usepackage{dcolumn}
\usepackage{bm}
\usepackage{color}
\usepackage{txfonts}

\begin{document}

\preprint{PRESAT-9601}

\title{Density functional theory study of effect of NO annealing on electronic structure and carrier-scattering property of 4H-SiC(0001)/SiO$_2$ interface}

\author{Nahoto Funaki}
\affiliation{Department of Electrical and Electronic Engineering, Graduate School of Engineering, Kobe University, Nada, Kobe 657-8501, Japan}
\author{Kosei Sugiyama}
\affiliation{Department of Electrical and Electronic Engineering, Graduate School of Engineering, Kobe University, Nada, Kobe 657-8501, Japan}
\author{Mitsuharu Uemoto}%
\affiliation{Department of Electrical and Electronic Engineering, Graduate School of Engineering, Kobe University, Nada, Kobe 657-8501, Japan}
\author{Tomoya Ono}
\affiliation{Department of Electrical and Electronic Engineering, Graduate School of Engineering, Kobe University, Nada, Kobe 657-8501, Japan}

\date{\today}

\begin{abstract}
The effect of the nitrided layers introduced by NO annealing on the electronic structure and carrier-scattering property of the 4H-SiC(0001)/SiO$_2$ interface is investigated by density functional theory calculations using the interface models where the areal N atom density corresponds to that in practical devices. It is found that the nitrided layer screens the unfavorable Coulomb interaction of the O atoms in the SiO$_2$. However, the electrons flowing under the nitrided layer are significantly scattered by the fluctuation of potential due to the low areal N atom density in practical devices, resulting in the low Hall mobility. These results imply that the areal N atom density should be increased so that the fluctuation of potential is suppressed.
\end{abstract}

\maketitle


SiC is a suitable material for future power electronics devices with high breakdown field, high carrier velocity, and a native oxide of SiO$_2$, making it ideal for metal-oxide-semiconductor field-effect transistors (MOSFETs). Among hundreds of polymorphs of SiC (e.g., 3C, 4H, and 6H), the most interesting commercially available polymorph is 4H-SiC, which can be grown as single-polymorph wafers.\cite{JpnJApplPhys_45_007565,JpnJApplPhys_54_040103,JpnJApplPhys_54_04DP07,ApplPhysExpress_13_120101} However, the high channel resistance of SiC-MOSFETs limits their performance.\cite{PhysStatusSolidiA_162_321,ApplPhysLett_77_003281} This high resistance is expected to be attributed to the low field-effect mobility in SiC-MOSFETs, which is much lower than the ideal electron mobility of a SiC bulk ($\sim$ 1000 cm$^2$V$^{-1}$s$^{-1}$).\cite{MaterSciForum_433-436_443} By post-oxidation annealing with nitric oxide (NO),\cite{IEEEElecDevLett_22_000176,ApplPhysLett_70_002028} the maximum field-effect mobility increases from 1--7 cm$^2$V$^{-1}$s$^{-1}$\cite{JApplPhys_91_001568} to 25--40 cm$^2$V$^{-1}$s$^{-1}$ for a 4H-SiC(0001) MOSFET.\cite{IEEETransElectronDevices_60_001260, IEEETransElectronDevices_62_000309,ApplPhysExpress_10_046601} The field-effect mobility is proportional to the product of Hall mobility and mobile carrier density. Hatakeyama {\it et al.} reported that the mobile carrier density is increased by NO annealing while the Hall mobility is not improved and the improvement in the field-effect mobility is due to the increase in the mobile carrier density.\cite{ApplPhysExpress_12_021003} However, the effect of the nitrided layers introduced by NO annealing on the low Hall mobility and the increase of mobile carrier density is not fully clear.

Most SiC-MOSFETs are $n$-type, with electrons in the conduction band serving as carriers. The behavior of conduction band edge (CBE) states of a SiC bulk is like that of free electrons, which are so-called ``floating states,''\cite{PhysRevLett_108_246404} and sensitive to the local atomic structure of interfaces.\cite{JPhysSocJpn_85_024701} In practical devices, the 4H-SiC(0001) surface is not atomically flat, e.g., the 4H-SiC(0001) substrates possess atomic scale steps on their surfaces which are generated during fabrication processes.\cite{SSDM_1987_227} In our previous study, we have reported that the CBE states are affected by the unfavorable Coulomb interaction of the O atoms in the SiO$_2$, which results in the discontinuity of the inversion layers at the step edge under a gate bias.\cite{ApplPhysExpress_17_011009} After NO annealing is carried out, the inserted nitrided layer screens the Coulomb interaction of the O atoms and the inversion layer becomes spatially continuous, resulting in the increase of mobile carrier density. However, in the computational model for the interface after NO annealing in our previous study, the areal N atom density was the same as the areal C atom density in a SiC bilayer and three times higher than that of practical devices.\cite{JSurfSciNanotechnol_15_109} In this paper, density functional theory (DFT)\cite{PhysRev_136_B864} calculations are carried out to investigate the effect of the nitrided layers in which the areal N atom density is reduced so as to correspond to that in practical devices. It is found that the nitrided layer screens the Coulomb interaction even when the areal N atom density is reduced. However, the conducting electrons from the source to the drain are scattered by the fluctuation of potential due to the low areal N atom density, which implies that the interface with the high areal N atom density is advantageous for electron conduction. 

\begin{figure}
\includegraphics{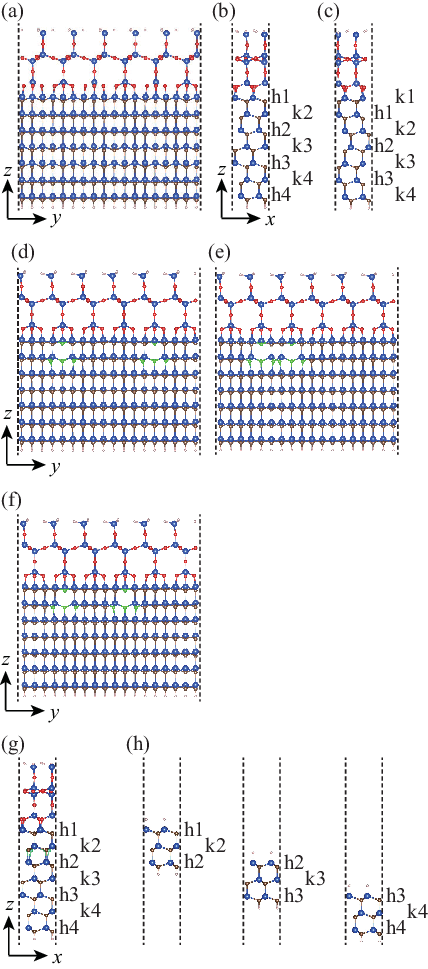}
\caption{Computational models for SiC(0001)/SiO$_2$ interface and SiC thin film. (a) Front view of the interface without any nitrided regions, (b) side view of the h-type interface without any nitrided regions, (c) side view of the k-type interface without any nitrided regions, (d)--(f) front views of the interface with nitrided regions, (g) side view of the h-type interface with nitrided regions, and (h) side views of thin films for the h-type interface. (d)--(f) are labeled models 1, 2, and 3, respectively. Blue, red, brown, green, and gray balls are Si, O, C, N, and H atoms, respectively. Since the atomic structure of the k-type interface looks the same as that of the h-type interface when it is seen from the [11$\bar{2}$0] direction, only the atomic structures of the h-type interfaces are shown for the front view. The number after h or k is the index of layer.}
\label{fig:1}
\end{figure}

\begin{figure}
\includegraphics{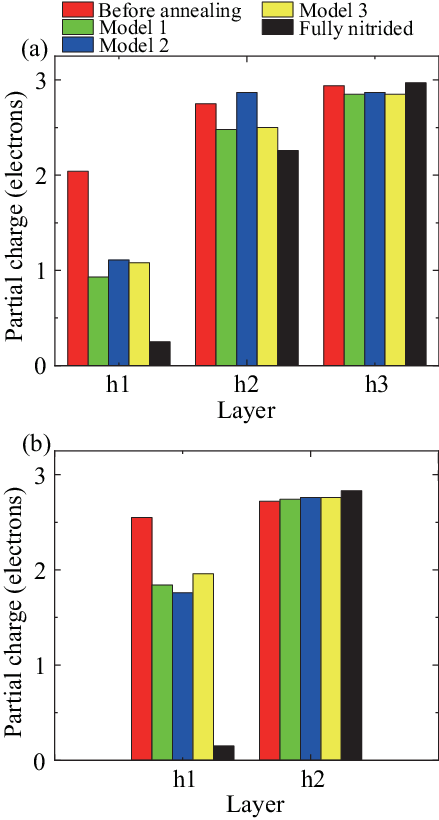}
\caption{Magnitude of partial charges of CBE states for (a) h- and (b) k-type interfaces in units of electrons. The indices of layer, h1, h2, and h3, correspond to those in Fig.~\ref{fig:1}.}
\label{fig:2}
\end{figure}

\begin{figure}
\includegraphics{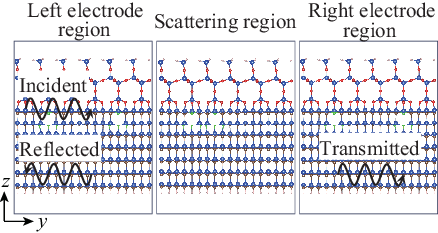}
\caption{Computational models for carrier-scattering property calculation. Blue, red, brown, green, and gray balls are Si, O, C, N, and H atoms, respectively.}
\label{fig:3}
\end{figure}

\begin{figure}
\includegraphics{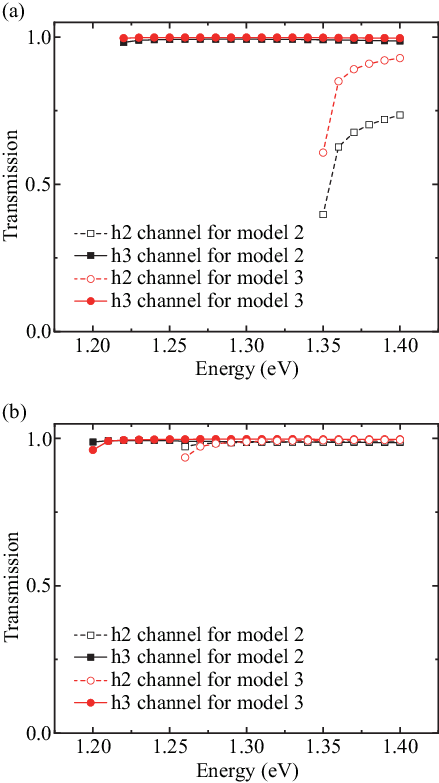}
\caption{Transmissions of h2 and h3 channels of (a) h- and (b) k-type interfaces. Energy is measured from the Fermi level.}
\label{fig:4}
\end{figure}

\begin{figure}
\includegraphics{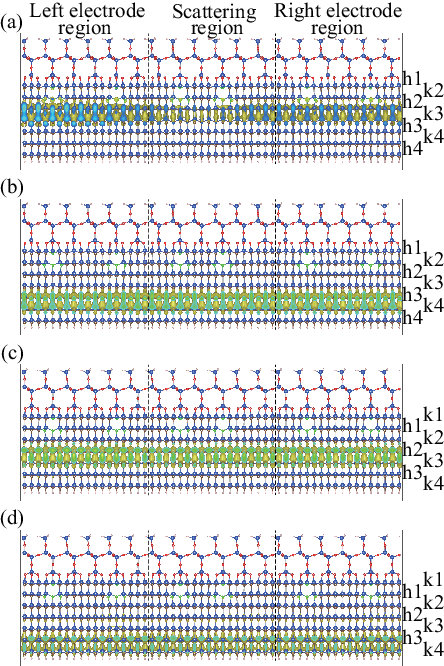}
\caption{Charge density distributions of scattering wavefunction associated with (a) h2 channel of h-type interface, (b) h3 channel of h-type interface, (c) h2 channel of k-type interface, and (d) h3 channel of k-type interface for incident electrons with energy of $\varepsilon_F$+1.38 eV. Blue, red, brown, green, and gray balls are Si, O, C, N, and H atoms, respectively. The indices of layer correspond to those in Fig.~\ref{fig:1}.}
\label{fig:5}
\end{figure}

The computational models for the electronic structure calculation are shown in Figs.~\ref{fig:1}(a) -- \ref{fig:1}(g). The RSPACE code\cite{PhysRevLett_82_005016,KikujiHirose2005,PhysRevB_82_205115} which uses the real-space finite-difference approach\cite{PhysRevLett_72_001240,PhysRevB_50_011355} for the DFT\cite{PhysRev_136_B864} is employed. According to scanning transmission electron microscopy images,\cite{PhysRevB_96_115311} most of the SiO$_2$ in the SiC(0001)/SiO$_2$ interface is amorphous. However, it is not straightforward to characterize the interface atomic structure. The crystalline interface atomic structures that can exist locally at the SiC(0001)/SiO$_2$ interface proposed in our previous study\cite{PhysRevB_96_115311} are employed. Our model contains a crystalline substrate with seven SiC bilayers (17.5 \AA~thick) connected without any coordination defects to a crystalline SiO$_2$ with a thickness of 8.2 \AA. In a 4H-SiC bulk, SiC bilayers are stacked on the quasi-cubic (k) and hexagonal (h) sites alternately along the [0001] direction. We refer to the interface where the Si atoms in the topmost SiC bilayer are on the k (h) site as the k (h)-type interface. The electronic structures of the k-type interfaces are different from those of the h-type interfaces.\cite{JPhysSocJpn_85_024701} Although it remains to be clarified whether the inserted N atoms exist on the SiO$_2$ or SiC side of the interface, we assume that the N atoms accumulate on the SiC side and the nitrided layers are formed at the interface on the basis of the possibility proposed in the previous experimental studies.\cite{JSurfSciNanotechnol_15_109,JApplPhys_97_074902, ApplPhysExpress_11_101303,PhysRevLett_98_136105,PhysRevB_79_241301} For the nitrided region, we adopt the models employed in our previous study, where a Si atom is removed and four C atoms with dangling bonds (DBs) are replaced by N atoms.\cite{JPhysSocJpn_90_124713} This structure has been proposed by Shirasawa {\it et al.}\cite{PhysRevLett_98_136105,PhysRevB_79_241301} using low-energy electron diffraction analysis and a couple of N atoms accumulate near the Si vacancy to terminate the DBs of a Si atom in this structure. When all the possible sites of the interface are replaced by the nitrided regions, the areal N atom density is 1.22 $\times 10^{15}$ atom/cm$^2$, which is three times higher than that in the practical device.\cite{JSurfSciNanotechnol_15_109} We employ the interface models in which one-third of the possible sites are replaced by the nitrided regions. Hereafter, the nitrided layer with the high (low) areal N atom density is referred to as the fully (partially) nitrided layer. Three types of interface model are prepared as shown in Figs.~\ref{fig:1}(d) -- \ref{fig:1}(f) according to the spatial distribution of the nitrided regions. The DBs at the top surface of the SiO$_2$ and the bottom surface of the SiC substrate are terminated by H atoms. The grid spacing in real space is taken to be 0.18 $\times$ 0.19 $\times$ 0.18 \AA$^3$. The exchange-correlation interaction is treated within the local density approximation of the DFT.\cite{CanJPhys_58_001200} For electronic structure calculations, the periodic boundary condition is imposed on all the directions. The supercell size is 5.33 $\times$ 27.72 $\times$ 40.44 \AA$^3$. Integration over the Brillouin zone is performed using a 6 $\times$ 2 $\times$ 1 k-point grid including $\Gamma$ point. The projector augmented wave method\cite{PhysRevB_50_017953} is adopted to describe the electron-ion interaction. We implement structural optimization until all the force components decrease to below 0.05 eV/\AA, while the atomic coordinates of the SiC bilayer in the bottom layer and the H atoms terminating DBs are fixed during the structural optimization.

The distribution of the charges associated with the CBE states is investigated using the partial charges,\cite{ApplPhysExpress_17_011009} in which the wavefunctions of the CBE states of the interface models are projected on those of the thin-film models as
\begin{widetext}
\begin{equation}
\rho_{PC}=\Sigma_{i,k} \left| \langle \psi_{i,k} | \phi_k \rangle \right|^2 \theta(\epsilon_{i,k}-\epsilon_F) \theta(\epsilon _{max}-\epsilon_{i,k}) \Delta_k,
\label{eqn:1}
\end{equation}
\end{widetext}
where $\epsilon_F$ is the Fermi level, $\theta$ is the Heaviside function, and $\phi_k$ is the wavefunction of the CBE states obtained by the thin-film models. $\epsilon_{max}$(=$\epsilon_F$+1.65 eV), which is the maximum energy of the energy window, is chosen so that the energy window contains the floating states inside the SiC substrate. The thin-film models for obtaining the partial charges are illustrated in Fig.~\ref{fig:1}(f). SiC bilayers are stacked with {\it hkh} ordering in the thin-film model because it is reported that the CBE states lie below the Si atom of the h site in a 4H-SiC(0001) substrate.\cite{JPhysSocJpn_85_024701} Therefore, for example, the partial charge of the h1 layer is obtained when the wavefunctions of the interface model are projected on those computed by the thin-film model shown in the left panel of Fig.~\ref{fig:1}(f). The DBs on both sides of the surface of the thin films are terminated by H atoms. Although the termination of surface atoms may change the magnitude of the partial charges, our conclusion is not affected because the same termination is employed for all thin-film models of SiC. The supercell size is chosen to be 5.33 $\times$ 27.72 $\times$ 40.44 \AA$^3$, which is the same as those for the interface models.

The partial charges of the interface with the nitrided regions are shown in Fig.~\ref{fig:2}. The magnitude of the partial charge in the partially nitrided layers is insensitive to the spatial distribution of the nitrided regions while it is smaller than that in the layers without the nitrided regions. The CBE states at the interface are removed even when the areal N atom density is reduced and the Coulomb interaction of the O atoms in the SiO$_2$ is screened by the partially nitrided layer. However, the magnitude of the partial charge in the partially nitrided layer is larger than that in the fully nitrided layer, indicating that the effect of NO annealing may be limited. In addition, the unfavorable Coulomb interaction is more severe for the h-type interface than the k-type interface.\cite{ApplPhysExpress_17_011009,PhysRevB_96_115311,PhysRevB_95_041302} The large reduction of the partial charge of the h-type interface indicates that the partially nitrided layer in the h-type interface can contribute to increase the mobile carrier density.

We then investigate the carrier-scattering property at the interface because the Kohn-Sham effective potential is disturbed by the nonuniformly inserted nitrided regions. Figure~\ref{fig:3} shows the computational models for the carrier-scattering property calculation. The computational models are divided into three parts; left electrode, scattering, and right electrode regions and the Kohn-Sham effective potential of these regions is obtained by imposing a periodic boundary condition in the $y$ direction. The left and right electrodes are set to be model 1 in Fig.~\ref{fig:1}(d) where the nitrided regions are inserted at equal intervals and the scattering region is set to be model 2 in Fig.~\ref{fig:1}(e) or model 3 in Fig.~\ref{fig:1}(f) so that the intervals of the nitrided regions are disordered. The electron-ion interaction is described by the norm-conserving pseudopotentials proposed by Troullier and Martins.\cite{ComputMaterSci_14_000072,PhysRevB_43_001993,PhysRevLett_48_001425} The scattering wavefunctions for the incident electrons from the left electrode to the right electrode are evaluated by the overbridging-boundary matching method.\cite{PhysRevB_67_195315,PhysRevB_70_033403,PhysRevB_98_195422}

The transmissions of the h- and k-type interfaces are shown in Fig.~\ref{fig:4}. The charge density distributions of the scattering wavefunction are plotted in Fig.~\ref{fig:5}. The transmission channels can be categorized into two types from the charge density distribution of the scattering wavefunction. The channels where the electrons accumulate near the interface and inside the substrate are referred to as the h2 and h3 channels, respectively. It is found that the transmissions of the h2 channel are lower than those of the h3 channel. In addition, the electrons flowing in the h2 channel of the h-type interface are more significantly scattered than those flowing in the h2 channel of the k-type interface because the h2 channel of the h-type interface is closer to the nitrided layer than that of the k-type interface. These results indicate that the fluctuation of potential causes the carrier scattering. In our previous study, we have reported that the O interstitials, which appear during thermal oxidation, significantly scatter electrons as well as the typical interface defects at the SiC/SiO$_2$ interface.\cite{PhysRevB_95_041302} The present results for the partially nitrided layer are consistent with the conclusion of our previous study on O interstitials. In addition, the Hall mobility remains low even after NO annealing as reported by Hatakeyama {\it et al.}\cite{ApplPhysExpress_12_021003} and the N atoms incorporated into the SiC side of the interface do not contribute to improve Hall mobility as reported by Hosoi {\it et al.}\cite{ApplPhysExpress_15_061003} The significant carrier scattering at the partially nitrided layer agrees with their experimental results. In the nitrided region proposed by low-energy electron diffraction analysis,\cite{PhysRevLett_98_136105,PhysRevB_79_241301} there are a couple of N atoms in the vicinity of the Si vacancy to passivate the DBs and the potentials of the partially nitrided layer, where the areal N atom density is low, are fluctuated. Concerning the carrier-scattering property, the present results suggest that the fully nitrided layer with the high areal N atom density is better than the partially nitrided layer.

In summary, the effect of the nitrided layer introduced by NO annealing on the electronic structure and carrier-scattering property of the SiC(0001)/SiO$_2$ interface has been investigated by DFT calculations. The areal N atom density was reduced from our previous study to replicate the interface of practical devices after NO annealing. It is found that the unfavorable Coulomb effect of the O atoms in the SiO$_2$ can be screened by the nitrided layer even when the areal N atom density is reduced. However, the conducting electrons flowing just beneath the partially nitrided layer are scattered in the case of the h-type interface. On the other hand, in the case of the k-type interface, the transmission of the conducting electrons is unity because the layer where the electrons flow is separated from the nitrided layer. The fluctuation of potential due to the low areal N atom density reduces the transmission and the scattering mechanism is the same as that at the interface after thermal oxidation,\cite{PhysRevB_95_041302} suggesting that the fully nitrided layer with the high areal N atom density is better than the partially nitrided layer from the view point of carrier-scattering property.

This work was partially financially supported by MEXT as part of the JSPS KAKENHI (24H01196, 24K01346). The numerical calculations were carried out using the computer facilities of the Institute for Solid State Physics at The University of Tokyo, the Center for Computational Sciences at University of Tsukuba, and the supercomputer Fugaku provided by the RIKEN Center for Computational Science (Project ID: hp230175, hp240178, hp250193).

\section*{Data availability}
The data that support the findings of this study are available from the corresponding author upon reasonable request.

\appendix


\bibliography{main}
\bibliographystyle{apsrev4-1}

\end{document}